# Predicting and interpreting oxide glass properties by machine learning using large datasets

Daniel R. Cassar[1], Saulo Martiello Mastelini[2], Tiago Botari[2], Edesio Alcobaça[2], André C.P.L.F. de Carvalho[2], Edgar D. Zanotto[1]

[1]Department of Materials Engineering, Federal University of São Carlos, São Carlos, Brazil
[2]Institute of Mathematics and Computer Sciences, University of São Paulo, São Carlos, Brazil

## Abstract

With the advent of powerful computer simulation techniques, it is time to move from the widely used knowledge-guided empirical methods to approaches driven by data science, mainly machine learning algorithms. We investigated the predictive performance of three machine learning algorithms for six different glass properties. For such, we used an extensive dataset of about 150,000 oxide glasses, which was segmented into smaller datasets for each property investigated. Using the decision tree induction, k-nearest neighbors, and random forest algorithms, selected from a previous study of six algorithms, we induced predictive models for glass transition temperature, liquidus temperature, elastic modulus, thermal expansion coefficient, refractive index, and Abbe number. Moreover, each model was induced with default and tuned hyperparameter values. We demonstrate that, apart from the elastic modulus (which had the smallest training dataset), the induced predictive models for the other five properties yield a comparable uncertainty to the usual data spread. However, for glasses with extremely low or high values of these properties, the prediction uncertainty is significantly higher. Finally, as expected, glasses containing chemical elements that are poorly represented in the training set yielded higher prediction errors. The method developed here calls attention to the success and possible pitfalls of machine learning algorithms. The analysis of the SHAP values indicated the key elements that increase or decrease the value of the modeled properties. It also estimated the *maximum* possible increase or decrease. Insights gained by this analysis can help empirical compositional tuning and computer-aided inverse design of glass formulations.

# 1. Introduction

Ceramics, copper, iron, and glasses are among the oldest synthetic, human-made materials. Objects made of glass have been known and extensively used for over 6,000 years. Vitreous materials have been systematically studied in the past two centuries; searches on the Scopus database and Derwent Innovation Index indicate that over half a million scientific articles have already been published. A similarly high number of patents have been filed on vitreous materials [1]. The SciGlass database also shows that over 400,000 inorganic glass compositions have already been disclosed [2], whereas many others are still hidden in industrial laboratories.

Glasses have evolved considerably from domestic use to an increasing number of high-tech applications, such as optical fibers, ionic conducting materials, optically functional formulations, bioactive compositions, and mechanically strong glasses and glass-ceramics. Glasses became so relevant to humankind that some authors coined the current times as the "glass age" [3,4]. To celebrate these fascinating materials, the United Nations recently declared 2022 as the International Year of Glass.

By far, most of these thousands of glassmaking formulations resulted from empirical trial-and-error experimental approaches, which were mainly guided by experience and accumulated knowledge. However, with the advent of computer simulation techniques, the glass community moves to data analytics-based approaches. One of the most efficient strategies would be to profit from the plethora of available composition-property datasets for data-driven modeling using Artificial Intelligence, particularly machine learning (ML) algorithms. The ultimate goal of the ML-based strategies is to specify a set of desired properties and find candidate compositions. However, before this task is accomplished, the creation of predictive models by ML algorithms, extracting rich and novel knowledge from several thousand glass composition-property values, is a required step. This is precisely where this work fits.

ML algorithms applied to relatively large datasets were investigated in oxide glasses since the 2003 pioneer work of Dreyfus and Dreyfus to predict the liquidus temperature for oxide glass-forming liquids [5], which is not a "glass property" but critical for glass making. Brauer et al. in 2007 [6] and Echezarreta-López et al. in 2013 [7] published the second and third scientific communications at the interface between ML and oxide glass. These two communications were about bioactive glasses. This topic gained traction, and at least fifteen communications have been published since [8–23]. Different oxide glass properties were investigated, such as solubility/dissolution [6,11,18,21], antimicrobial [7], transmittance [8], liquidus temperature [9,13,19,23], Young's modulus [9,13–15,17,19,22,23], average fiber diameter [10], glass transition temperature [12,19,20,23], compressive stress [13], coefficient of thermal expansion [19,23], depth of ionic exchange layer [13], diffusion [13], viscosity [13,24], density [17,22,23], Poisson's ratio [17], sheer modulus [17,19,22,23], refractive index [19,23], and hardness [19]. Most of these properties and more were studied in a recent paper by Ravinder and co-authors [25].

Many ML algorithms and AI strategies were used in the works mentioned in the previous paragraph, such as neural networks (NNs) [6,8–15,17–21,24], neurofuzzy systems [7], elastic net/lasso/ridge/linear/polynomial/additive regression [11,13,14,17,18,21,22], random forest (RF) [11,14,17,18,20,21], support vector machines (SVMs) [11,17,18,20,21], Gaussian process regression [15,18,21,23], cluster analysis [12], k-nearest neighbor (k-NN) [17,20], classification and regression tree [18,20], boosting methods [18–20,22,25], and M5P [22]. As one would expect, the strategy, scope, dataset size, and prediction/generalization power vary significantly between all these published works.

NNs, while not the object of study of this particular work, are by far the most used ML algorithm for property prediction of oxide glasses (15 publications). This is probably due to the continuous and non-linear chemical dependence of many glass properties, a practical problem to use NNs. Similarly, the most studied property in this context is the Young's modulus (8 publications), which is also one of the properties that we investigated in this work.

In general, published works observed mild to great success at using ML algorithms to predict oxide glass properties. In the following paragraphs, we focus on the six properties (glass transition temperature, liquidus temperature, elastic modulus, thermal expansion coefficient, refractive index, and the Abbe number) that we studied in this work and previously reported results.

*Glass transition temperature* was previously studied by us [12,20], and Ravinder et al. [19], and Bishnoi et al. [23]. In this work, we revisited the investigation reported in a previous communication [20] to compare with new results. Ravinder et al. [19] trained a NN on an undisclosed number of data points (by analyzing their Figure S3f, we estimate it was ca. 12,000) and obtained a value of 0.90 for the $R^2$ metric when predicting the test dataset. Additionally, Bishnoi et al. [23], using Gaussian processes regression (GPR), trained a model over the same database of [19], obtaining a value of 0.90 for $R^2$, and Ravinder et al. [25], using XGBoost, obtained a value of 0.94. Here we studied approximately 50,000 glasses with different chemical compositions and used three other ML algorithms not considered in refs. [19,23], and obtained an $R^2$ value of 0.96 for our best model (RF) in the test dataset.

*Liquidus temperature* was previously studied by Dreyfus et al. [5], Mauro et al. [9], Tandia et al. [13], Ravinder et al. [19], and Bishnoi et al. [23]. Except for the last, all of them trained NNs to predict this property. The first two investigated a limited dataset with no more than 1,000 examples, using a much more restricted compositional set than we studied here. Both Tandia et al. and Ravinder et al. investigated an undisclosed number of data points (we estimate it to be about 4,000 for the first by analyzing their Figure 33.10, and about 7,000 for the second by analyzing their Figure S3g). Tandia et al. proposed a clustering analysis after observing that studying all data points together yielded some predictions with low accuracy. For their largest cluster, they obtained a value of 0.89 of $R^2$ for the test set. A smaller value of 0.80 was obtained by Ravinder et al. for their test set. Bishnoi et al. [23], using the same dataset as Ravinder et al. [19], trained a GPR model, obtaining a value of 0.85 for $R^2$. Ravinder et al. [25], using XGBoost, obtained a value of 0.95 for $R^2$. Here we studied about 33,000 glasses with different chemical compositions and used three other ML algorithms not considered in refs. [5,9,13,19], and obtained an $R^2$ value of 0.98 for our best model (RF) in the test dataset.

The *refractive index* was previously studied by Ravinder et al. [19] and Bishnoi et al. [23], where they trained an NN and a GPR, respectively, on an undisclosed number of data points (we estimate it was about 18,000 by analyzing Figure S3h in ref. [19]). Ravinder et al. [19] obtained an $R^2$ of 0.94, whereas Bishnoi et al. [23] obtained an $R^2$ of 0.96 for the GPR model, both calculated over the test dataset. Ravinder et al. [25] obtained a value of 0.94 using XGBoost, and Zaki et al. [26] obtained a value of 0.97 using NNs. Here we studied about 45,000 glasses with different chemical compositions and used three other ML algorithms not considered in refs. [19,23], and obtained an $R^2$ value of 0.97 for our best model (RF) in the test dataset.

*Young's modulus* is the most studied property in the context of ML applied to oxide glasses [9,13–15,17,19,22,23]. Many published works focus on limited datasets with no more than 1,000 data points or study synthetic data created by atomistic simulations. Deng [17] studied about 25,000 data points from a proprietary Corning database and tested four ML algorithms: k-nearest neighbor, NNs, SVMs, and lasso linear regression. His best result was for the first three algorithms, for which he obtained an $R^2$ of 0.96. Ravinder et al. [19] trained a NN on an undisclosed number of data points (we estimate it was about 7,000 by analyzing their Figure S3b) and obtained a value of 0.86 for the $R^2$ when predicting the test dataset. Using the same dataset as Ravinder et al. [19], Bishnoi et al. [23] trained a GPR model and obtained an $R^2$ of 0.90. Ravinder et al. [25], using XGBoost, obtained a value of 0.90 for $R^2$. Here we studied about 13,000 glasses with different chemical compositions, used a regression tree for the first time to predict this property, and obtained an $R^2$ value of 0.92 for our best model (RF) in the test dataset.

The *coefficient of thermal expansion* was previously studied by Ravinder et al. [19] and Bishnoi et al. [23], whom both used the same dataset on an undisclosed number of data points (we estimate it was about 18,000 by analyzing their Figure S3e). Ravinder et al. [19] used a NN and obtained an $R^2$ value of 0.80, while Bishnoi et al. [23] used a GPR model and obtained an $R^2$ of 0.83, both predicting over test dataset. Ravinder et al. [25], using XGBoost, obtained a value of 0.89 for $R^2$. We studied about 51,000 glasses with different chemical compositions, using three other ML algorithms not considered in refs. [19,23], and obtained an $R^2$ value of 0.94 for our best model (RF) in the test dataset.

While the preprint of this communication was, to the best of our knowledge, the first to report ML algorithms to predict the *Abbe number* of glasses, since then, two other works also reported similar models; Ravinder et al. [25] obtained a value of 0.97 for $R^2$ using XGBoost, while Zaki et al. [26] obtained a value of 0.95 using NNs. Our best model for predicting the Abbe number was RF, and we obtained an $R^2$ value of 0.98 over the test dataset.

Finally, a challenge posed to the glass community [27] refers to the interpretation of the ML-induced models, that is, extracting meaningful information from the trained ML models that can give new insights. A novel model-agnostic approach that can be helpful in this front is the computation of the SHAP (SHapley Additive exPlanations) values [28]. SHAP values are based on the Shapley values from coalitional game theory [29] and explain the prediction of models by computing the contribution of each feature relative to a baseline predicted by a null model (the mean value of the target). In this work, the features are the quantities of chemical elements present in glasses, and the targets are one of the six properties. With the SHAP analysis, we can show the contribution of each element to the prediction of the property of interest, information that can significantly help in designing new glasses.

To the best of our knowledge, the first research work to use SHAP analysis in the context of oxide glasses is a preprint recently deposited by Zaki et al. [26]. In a follow-up preprint [25], the same group developed composition–property models for 25 properties. These models were interpreted using SHAP to understand the role of glass components in controlling any desired property. Their analysis revealed that the glass network formers, modifiers, and intermediates, play distinct roles in governing each property. Most interesting was their finding that these components exhibit *interdependence*, the magnitude of which is different for each property. The physical origins of some of these interdependencies could be attributed to known phenomena such as the boron anomaly, the mixed modifier effect, and the Loewenstein rule; however, some remaining ones require further experimental and computational analysis of the glass structure.

## 2. Objectives

In a previous work [20], we tested the ability of six ML algorithms to predict the glass transition temperature ($T_g$) of oxide glasses. Three of them were among the most successful and can also, in principle, bring valuable interpretations. They are:

- a decision tree induction algorithm (Classification And Regression Tree, CART);
- a lazy learning algorithm (k-Nearest Neighbors, k-NN);
- and an ensemble of tree-based regressors (Random Forest, RF).

However, $T_g$ is only one among several relevant properties of glasses. In this study, we test the predictive power of these three algorithms for $T_g$ and an additional set of five other important properties: liquidus temperature ($T_l$), Young's modulus (E), thermal expansion coefficient (CTE), refractive index ($n_D$), and Abbe number ($\upsilon_D$). In this study, we compare different ML algorithms and check which is most suitable to predict each investigated property.

## 3. Materials and methods

### 3.1 Machine learning algorithms

In this investigation, we used the three ML algorithms that presented the best predictive performance in our previous study [20], CART, k-NN, and RF.

The CART algorithm induces a decision tree for classification and regression tasks [30] in what is called a greedy training phase. While in classification tasks, the decision tree is named classification tree; in regression tasks, it is called a regression tree. The name "decision trees" is due to the upside-down tree shape of the induced models, when the leaf nodes are associated with a class (classification tree) or an actual value (regression tree), and each internal node is associated with a predictive attribute in the training set. In a regression tree (the decision tree induced in this study), each path from the tree root to a leaf node is a regression rule. A new instance is labeled by following a particular path, according to its predictive attribute values.

The k-NN algorithm is one of the simplest ML algorithms [31]. This algorithm is instance-based and does not have an explicit training phase. It builds a model only when a new instance needs to be labeled, known as lazy learning. In classification tasks, k-NN labels a new (unlabeled) instance with the most common label in the k closest training set instances. In the regression task, k-NN labels a new instance with the mean label value of the k closest training set instances. The components of the mean label calculation might be weighted according to their distance to the query instance.

The RF algorithm, instead of producing a single decision tree (as done by CART), induces a set of decision trees, creating a forest [32]. RF can also be used for classification and regression tasks. When used in regression tasks, each tree in the forest is induced by a dataset obtained by sampling (with replacement) the training set and randomly selecting a subset of the input features as candidate attributes for each split in the inner nodes of the regression trees. The process of inducing regression trees is based on the Bagging ensemble building procedure [33]. When applied to a new, unlabeled instance, in regression tasks, RF returns the average of the predictions from its regression trees.

We trained the models with large datasets, some with approximately 50,000 examples. Knowing that the performance of predictive models induced by ML algorithms can be affected by the values of their hyperparameters, we tuned the algorithms using Random Search [34] to reduce the RRMSE (relative root mean squared error) value. This technique is simple but effective, especially when not all hyper-parameters are equally important [34]. To do this, the dataset was split using 10-fold cross-validation in an outer loop and 5-fold cross-validation in an inner loop, using the same strategy employed in a previous communication of ours [20]. The hyperparameter space used for each ML algorithm is reported in the Supplementary Materials, alongside the best hyperparameter values found. We provide the codes and datasets used here on GitHub (https://github.com/ealcobaca/mlglass) to make the experiment fully reproducible. All calculations were done in a Python environment, using algorithms provided by the scikit-learn package [35], one of the most used modules by the machine learning community.

## 3.2 Data collection

We collected all the data used in the experiments from the SciGlass database [2]. As already mentioned, here we collected the following properties: glass transition temperature, liquidus temperature, Young´s modulus, thermal expansion coefficient, refractive index, and the Abbe number. Afterward, we performed a systematic data preparation and cleaning. The main steps are described in the following paragraphs.

We limited our investigation to oxide glasses, which we defined as glasses with an atomic oxygen fraction of at least 0.3. Moreover, we excluded glass compositions that contain the following elements: S, H, C, Pt, Au, F, Cl, N, Br, and I. The rationale behind this selection is that these elements only appear in very few compositions or are considerably volatile. Hence, their reported nominal contents are rarely close to the actual content, introducing unwanted errors during the models' training. Furthermore, some of these elements can occupy the atomic sites of oxygen in the glass structure, which we wanted to avoid.

Additionally, we removed the (incorrect) negative values of CTE and performed a transformation to the base-10 logarithm of this property. Unlike the other properties studied here (which typically vary by no more than 1 order of magnitude), the CTE varies by 2.5 orders of magnitude, from the lowest to the highest value of the dataset. We observed that the predictive performance of the ML models induced for this property was poor when we considered the CTE values on the linear scale (results not reported in this paper). To deal with this problem, we pre-processed this property applying the logarithm function in base 10.

We also eliminated the extreme values of each property. An extreme value is defined here as not within the range between 0.05% and 99.95% percentiles. From our experience, these glasses with extreme properties are especially prone to significant errors, including typos. Besides, the minimal number of instances with these extreme characteristics in the dataset renders their modeling very difficult.

We also eliminated the duplicate examples (glasses of the same nominal composition) by first rounding up the atomic fraction to the 3$^{rd}$ decimal place (i.e., three significant digits) and then grouping all examples by their chemical composition. The median of the property was taken from each composition group, and therefore the final dataset had only unique compositions. This strategy was successfully used in our previous work [20] and has the advantage of preventing data leakage [36].

The final dataset after all these steps can be found in the GitHub repository at https://github.com/ealcobaca/mlglass.

## 3.3 SHAP analysis

As mentioned in the introduction, we computed the SHAP values to interpret the prediction of the induced models. These values were estimated using the *shap* Python module [37]. SHAP values are computed for every feature of the models (quantity of each chemical element in this work) and have the same unit as the target property predicted by the models. Moreover, SHAP values are also additive, meaning that summing all the SHAP values for a given prediction plus a base value (which is the mean of the target value) will return the model prediction value. This additive property gives us the contribution of each feature to the final prediction, which, in this case, is relevant to the glass research community as it communicates *which* element contribute to increase or decrease a given property. Not only this, but we can also infer by *how much* each element increases or decreases the property.

There are many ways to visualize the SHAP values. In this work, we use the beeswarm plots. In these plots, the ten most relevant chemical elements are shown on the left, sorted by decreasing order of importance. The absolute sum of SHAP values measures "importance" in this context; the higher the importance, the bigger the impact of this element on the considered property. Each element has an associated horizontal axis with colored dots; each dot represents a glass from the studied dataset with the associated element. When dots have the same SHAP value, they stack vertically. The dot colors are related to the quantity of the respective chemical element in the glass, going from low to high, as shown by the color scale on the plot (see Figure 15). Finally, the *x*-axis value is the SHAP value that shows the impact of that element in the model prediction. A SHAP value of 5, for example, means that the addition of that quantity of the chemical element had an impact of increasing the property by 5 units from the base value.

All the beeswarm plots shown in this work were obtained using the induced CART model of the respective property and using all the available datasets. Computing the SHAP values for the induced RF models was not possible for all properties on the available hardware, where the major limiting factor was the RAM size (16 Gb).

## 4. Results and discussion

### 4.1 Analysis of the datasets used in this study

Tables 1 and Figures 1 and 2 summarize some statistical measures of the datasets for each investigated property, which span a range of at least 12,000 examples for Young's modulus, up to 51,000 for CTE. It is relevant to note that these datasets include glasses containing up to 23 different chemical elements. This rich data collection is by itself quite informative for the glass research community.

**Table 1**. Descriptive statistics of the used datasets.

|          | vD     | nD     | log10(CTE) | $T_{liq}$ (K) | $E$ (GPa) | $T_g$ (K) |
|----------|--------|--------|------------|---------------|-----------|-----------|
| Count    | 21,838 | 44,857 | 51,032     | 33,216        | 12,699    | 49,994    |
| Mean     | 43.0   | 1.69   | −5.12      | 1406          | 76.8      | 777       |
| Std      | 13.2   | 0.19   | 0.20       | 313           | 20.6      | 152       |
| Min      | 10.5   | 1.40   | −6.89      | 595           | 10.5      | 378       |
| 50%      | 42.4   | 1.64   | −5.09      | 1389          | 76.3      | 773       |
| Max      | 71.9   | 2.75   | −4.42      | 3061          | 162.6     | 1248      |
| Skewness | −0.06  | 1.39   | −1.02      | 1.03          | 0.22      | 0.15      |
| Kurtosis | −0.97  | 2.48   | 3.92       | 3.05          | 0.93      | −0.33     |

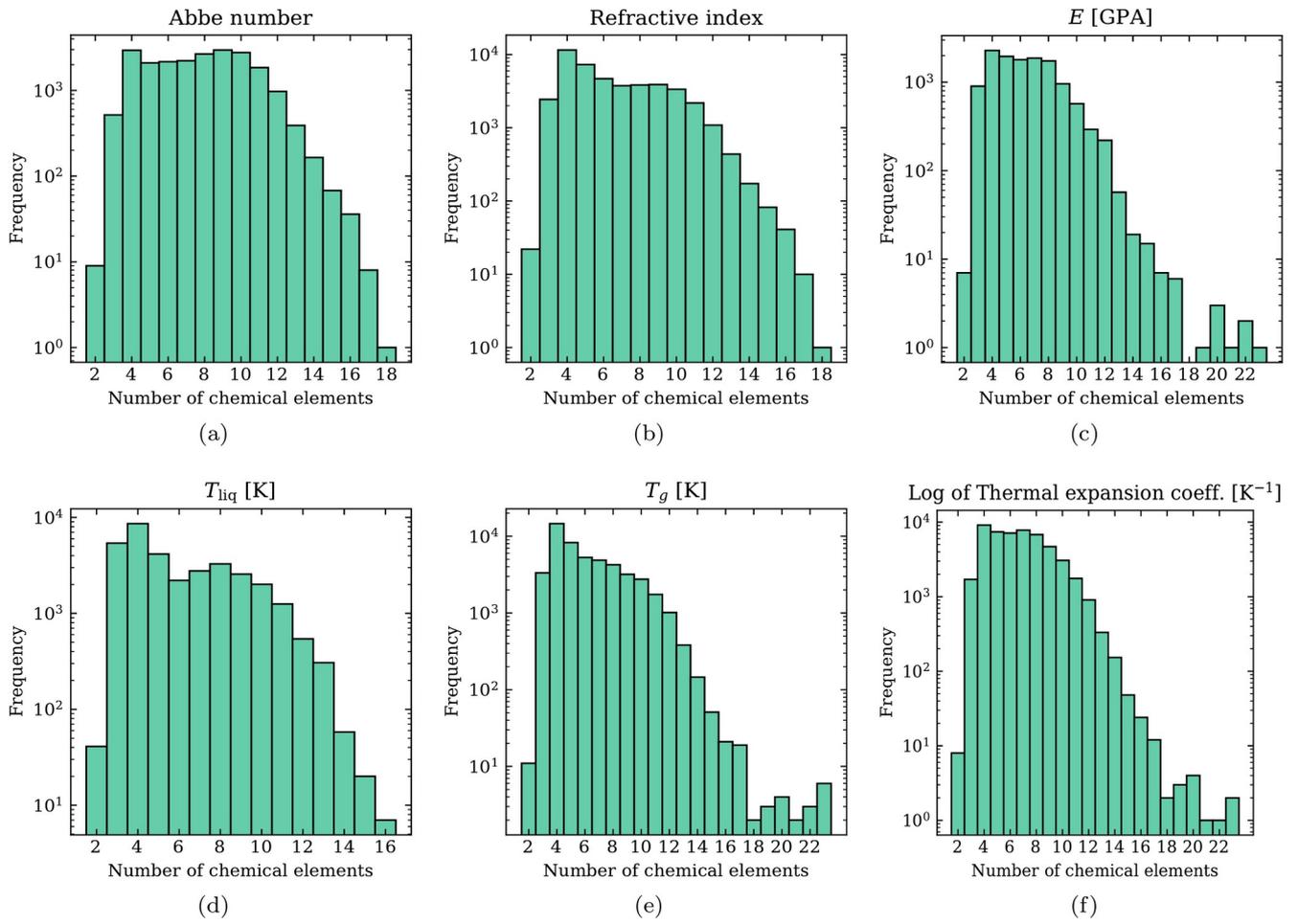

**Figure 1**. Frequency versus the number of chemical elements in each composition for six properties of oxide glasses.

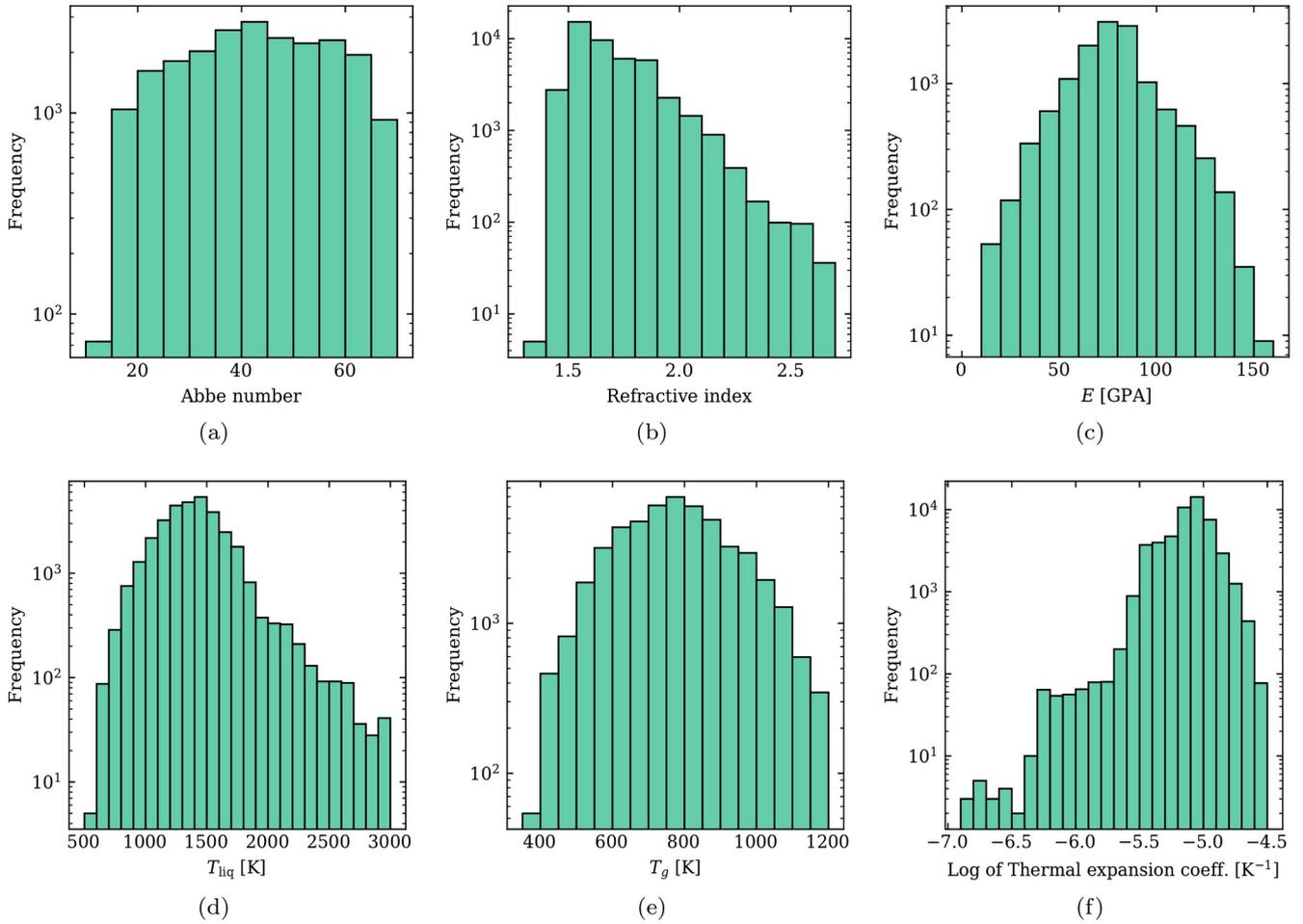

**Figure 2**. Frequency versus value for six properties of oxide glasses.

Figure 2 and the skewness values from Table 1 show that the distribution of values for some glass properties is more asymmetric than others. This is the case of $n_D$, $T_{liq}$, and $\log_{10}(CTE)$, with a significant right tail for the first two and a noticeable left tail for the last. While $\nu_D$, $T_g$, and $E$ distributions are reasonably symmetric and have a skewness close to zero.

## 4.2 Overview of experimental results

We carried out several ML experiments with and without hyperparameter tuning for the six glass properties investigated (see the section on Objectives for details). Tables 2, 3, and 4 show an overview of the metrics for the three ML algorithms that we considered (all in their tuned version). Overall, the RF yielded the lowest errors, closely followed by the k-NN, whereas the CART led to higher errors for these properties.

**Table 2**. Values of the performance metrics for all six properties obtained using the tuned CART algorithm. The upward arrows indicate that the higher the metric the better, whereas the downward arrows indicate the opposite.

| Metric | $v_D$ | $nD$ | $\log_{10}(CTE)$ | $T_{liq}$ (K) | $E$ (GPa) | $T_g$ (K) |
|---|---|---|---|---|---|---|
| RD (↓) | 3.67±0.15 | 1.13±0.05 | 0.80±0.02 | 2.98±0.08 | 7.0±0.5 | 3.40±0.09 |
| $R^2$ (↑) | 0.959±0.005 | 0.944±0.006 | 0.883±0.007 | 0.947±0.007 | 0.83±0.03 | 0.920±0.005 |
| RMSE (↓) | 2.8±0.2 | 0.044±0.002 | 0.0687±0.003 | 72±5 | 8.6±0.8 | 43±1 |
| RRMSE (↓) | 0.20±0.01 | 0.24±0.01 | 0.34±0.01 | 0.23±0.02 | 0.42±0.04 | 0.285±0.009 |

**Table 3**. Values of the performance metrics for all six properties obtained using the tuned k-NN algorithm. The upward arrow indicates that the higher the metric the better. The downward arrow indicates the opposite.

| Metric | $v_D$ | $nD$ | $\log_{10}(CTE)$ | $T_{liq}$ (K) | $E$ (GPa) | $T_g$ (K) |
|---|---|---|---|---|---|---|
| RD (↓) | 3.6±0.2 | 0.80±0.04 | 0.59±0.02 | 2.17±0.07 | 5.2±0.4 | 2.54±0.11 |
| $R^2$ (↑) | 0.960±0.004 | 0.965±0.005 | 0.931±0.006 | 0.974±0.002 | 0.90±0.02 | 0.953±0.004 |
| RMSE (↓) | 2.8±0.2 | 0.0345±0.003 | 0.0524±0.002 | 51.0±1.5 | 6.4±0.6 | 33±1 |
| RRMSE (↓) | 0.20±0.01 | 0.187±0.015 | 0.26±0.01 | 0.163±0.005 | 0.31±0.03 | 0.216±0.009 |

**Table 4**. Values of the performance metrics for all six properties obtained using the tuned RF algorithm. The upward arrows indicate that the higher the metric the better. The downward arrows indicate the opposite.

| Metric | $v_D$ | $nD$ | $\log_{10}(CTE)$ | $T_{liq}$ (K) | $E$ (GPa) | $T_g$ (K) |
|---|---|---|---|---|---|---|
| RD (↓) | 2.7±0.1 | 0.76±0.02 | 0.538±0.009 | 2.27±0.06 | 5.4±0.4 | 2.36±0.05 |
| $R^2$ (↑) | 0.977±0.004 | 0.974±0.004 | 0.943±0.005 | 0.976±0.002 | 0.92±0.02 | 0.964±0.002 |
| RMSE (↓) | 2.1±0.2 | 0.030±0.002 | 0.048±0.002 | 50±2 | 6.0±0.6 | 29.4±0.9 |
| RRMSE (↓) | 0.15±0.01 | 0.17±0.01 | 0.241±0.009 | 0.159±0.006 | 0.29±0.03 | 0.193±0.006 |

## 4.3 Glass transition temperature

The glass transition temperature is one of the most important properties of glasses from both science and technology perspectives [38]. It is related to the non-equilibrium nature of the temporarily frozen glassy state, which leads to spontaneous structural relaxation. It also determines the annealing process during manufacturing to relieve residual stresses (which cause warping and spontaneous fracture of glass articles) and limit the maximum temperature of use for structural components that have a glassy phase.

Although results for $T_g$ were published in our previous article [20], we have now made minor changes to the data collection strategy, discussed in Section 3.2. Therefore, in this article, we reanalyzed this property. Table 5 shows the predictive performance assessed by different statistical metrics, described in detail elsewhere [20].

**Table 5**: Experimental results for $T_g$. The full dataset contains about 50,000 different glass compositions.

| Metric | Cart | | k-NN | | RF | |
|---|---|---|---|---|---|---|
| | Default | Tuning | Default | Tuning | Default | Tuning |
| RD | 3.25 ± 0.06 | 3.40 ± 0.09 | 2.74 ± 0.07 | 2.5 ± 0.1 | 2.46 ± 0.05 | 2.36 ± 0.05 |
| $R^2$ | 0.920 ± 0.004 | 0.920 ± 0.005 | 0.949 ± 0.003 | 0.953 ± 0.004 | 0.957 ± 0.003 | 0.964 ± 0.002 |
| RMSE | 43.7±0.9 | 43.36 ± 1.03 | 34.5 ± 1.0 | 33.0 ± 1.2 | 31.7 ± 0.9 | 29.4 ± 0.9 |
| RRMSE | 0.287 ± 0.008 | 0.285 ± 0.009 | 0.226 ± 0.007 | 0.216 ± 0.009 | 0.208 ± 0.006 | 0.193 ± 0.006 |

Table 5 shows that, on average, the overall top performer for $T_g$ property is the tuned RF algorithm, followed by the RF with default hyperparameter values. For each algorithm, the tuned version was usually superior to the default version.

Figures 3a, b, and c show the box plots describing the relative deviation (residual of prediction divided by the reported value) as a function of the measured value of $T_g$ for the tuned version of each ML algorithm. The boxes are bound by the first and third quartiles (Q1 and Q3) and have caps representing the standard deviation of the data. The notch of the boxes is the 95% confidence interval of the median, which is shown by the horizontal orange line. Similar box plots will be used to investigate the other properties throughout the manuscript.

We included these error box plots for each ML algorithm for a better insight into the prediction uncertainty as a function of the actual value of $T_g$. These figures show that the prediction errors are significantly larger for the smallest and largest values of $T_g$. The algorithm with the best performance for extreme values is CART, closely followed by k-NN. The typical relative deviation (RD) with the k-NN and RF (2.5%) is quite reasonable and is similar to the actual variance in the reported data. However, for the extreme values, it can reach 10%. This result indicates that care must be taken when predicting novel glassmaking compositions with extreme $T_g$ values.

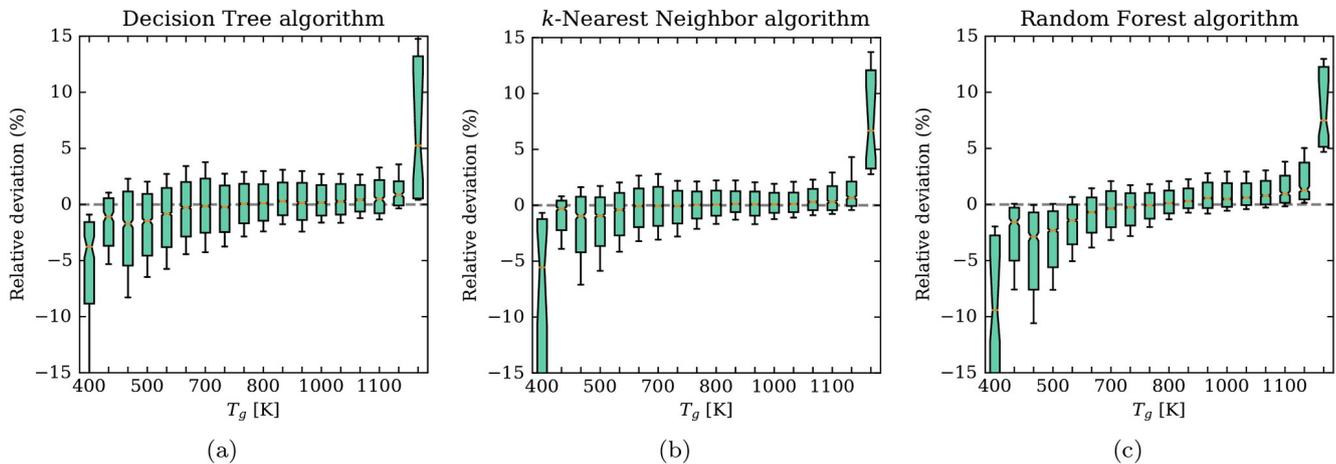

**Figure 3**. Boxplot of relative deviation (residual of prediction divided by the reported value) of $T_g$ for the tuned models.

Finally, Figure 4 shows the mean and standard deviation of the prediction residuals of $T_g$ for each chemical element (the residual is the difference between the reported and the predicted value). The model used to build this plot was the one with the best RD metric, the tuned RF. As one would expect, the standard deviations of the prediction residuals for the elements with fewer examples in the dataset (those in the far left of Figure 4) are systematically higher than those for the elements with more examples. Besides, the mean values of the residuals are systematically more spread for the elements with fewer examples than those with more examples. With only one example in this dataset, mercury had the highest residual, of about 100 K. In a previous communication, we opted to eliminate the chemical elements that appear in less than 1% of the examples [12]. The present result gives support to this strategy, at least for this particular property.

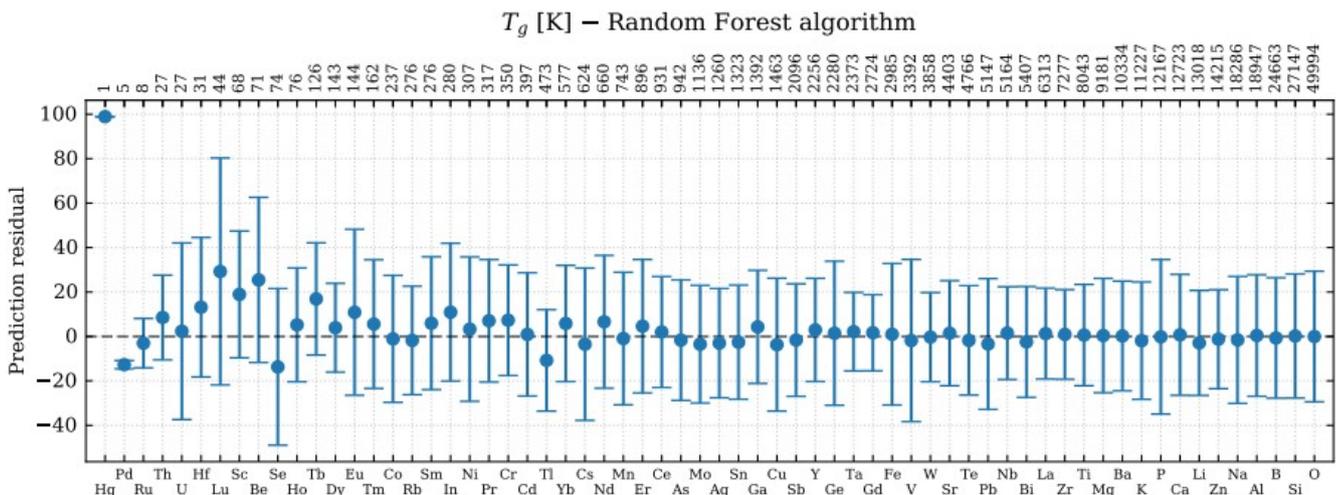

**Figure 4**. Mean and standard deviation of the prediction residual of $T_g$ for each chemical element. The numbers on top are the number of examples (glass compositions) containing that element in the dataset. The prediction residual is the difference between the reported and the predicted value.

It is relevant to note in this figure that the mean prediction residual (central dots) fluctuates significantly for the first 25 chemical elements, then consistently approaches the value zero when the number of examples reaches N > 600. Thus, when a particular element is present in 600 or more compositions, the prediction error of $T_g$ is minimized, on average.

## 4.5 Liquidus temperature

Knowledge of the liquidus temperature is essential for manufacturing glasses via melt and quench, which have to be melted in adequate inert, refractory materials, chemically homogenized and refined (eliminate bubbles) at temperatures above $T_{liq}$ [39]. However, it is well known by glass technology experts that experimental determination of the liquidus is challenging, highly time-consuming, and subject to substantial errors. Therefore, the capacity of predicting this property would be very relevant to the research and industrial community.

**Table 6**: Experimental results for $T_{liq}$. The full dataset contains about 33,000 different glass compositions.

| Metric | Cart | | k-NN | | RF | |
|---|---|---|---|---|---|---|
| | Default | Tuning | Default | Tuning | Default | Tuning |
| RD | 2.96 ± 0.07 | 2.98 ± 0.08 | 2.41 ± 0.06 | 2.17 ± 0.07 | 2.26 ± 0.07 | 2.27 ± 0.06 |
| $R^2$ | 0.945 ± 0.009 | 0.947 ± 0.007 | 0.969 ± 0.002 | 0.973 ± 0.002 | 0.971 ± 0.004 | 0.976 ± 0.002 |
| RMSE | 73.3 ± 6.0 | 72.4 ± 5.2 | 55.3 ± 2.1 | 51.0 ± 1.5 | 53.3 ± 3.9 | 49.6 ± 2.3 |
| RRMSE | 0.23 ± 0.02 | 0.23 ± 0.02 | 0.177 ± 0.006 | 0.163 ± 0.005 | 0.17 ± 0.01 | 0.159 ± 0.006 |

Table 6 shows that the top performer algorithms for the $T_{liq}$ are the k-NN and RF. The CART algorithm obtained the lowest overall performance. Once more, the tuned version usually presented a better performance than the default version, as expected.

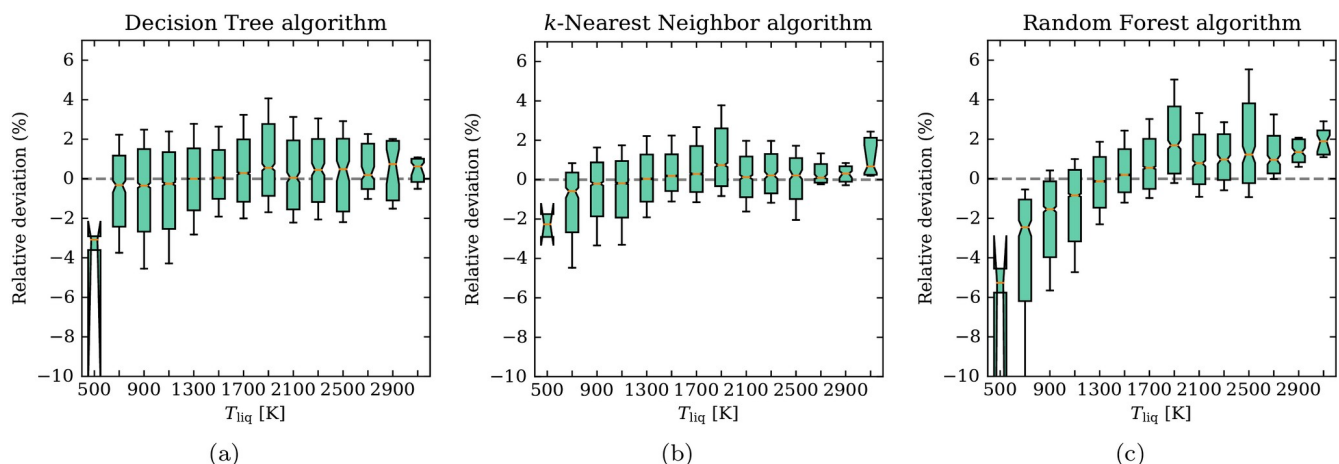

**Figure 5**. Boxplot of relative deviation (residual of prediction divided by the reported value) of $T_{liq}$ for the tuned models.

For a more in-depth insight into the prediction error as a function of the actual values of $T_{liq}$, we produced box plots for each tuned algorithm (Figures 5a, b, c). The prediction errors for k-NN show a very interesting trend: the errors in the extremes are higher, but not absurd, most of them between −3 and 3%. The tuned k-NN obtained the best overall result, approximately 2% for the RD measure. This average error of 2% is relatively low and is similar to the actual measurement error. This is an excellent result because, in principle, it predicts novel glassmaking compositions with either low, intermediate, or very high liquidus. However, as in the previous property, the uncertainty is substantially higher for extreme values. Due to the "lazy" nature of the k-NN algorithm, it is worth noting that it does not provide reliable predictions for properties outside the data distribution in the training set.

Finally, Figure 6 shows the mean and standard deviation of the prediction residuals of $T_{liq}$ for each chemical element. The model used to build this plot was the one with the lowest RD, the tuned k-NN. Three chemical elements worth mentioning are uranium, gallium, and cerium, as they are those with the highest standard deviation of the residuals. The first can be explained by the small number of examples in the dataset (only 7), but the other two have at least 186 examples. Further data and experiments are necessary to understand why their predictive error is higher than usual.

Similar to what we observed for $T_g$ (Figure 4), the mean values of the residuals are systematically more spread for the elements with fewer examples than for those with more examples. The mean prediction residuals significantly fluctuate for the first chemical elements, then remain constant and approach the experimental value for a number of examples N > 400.

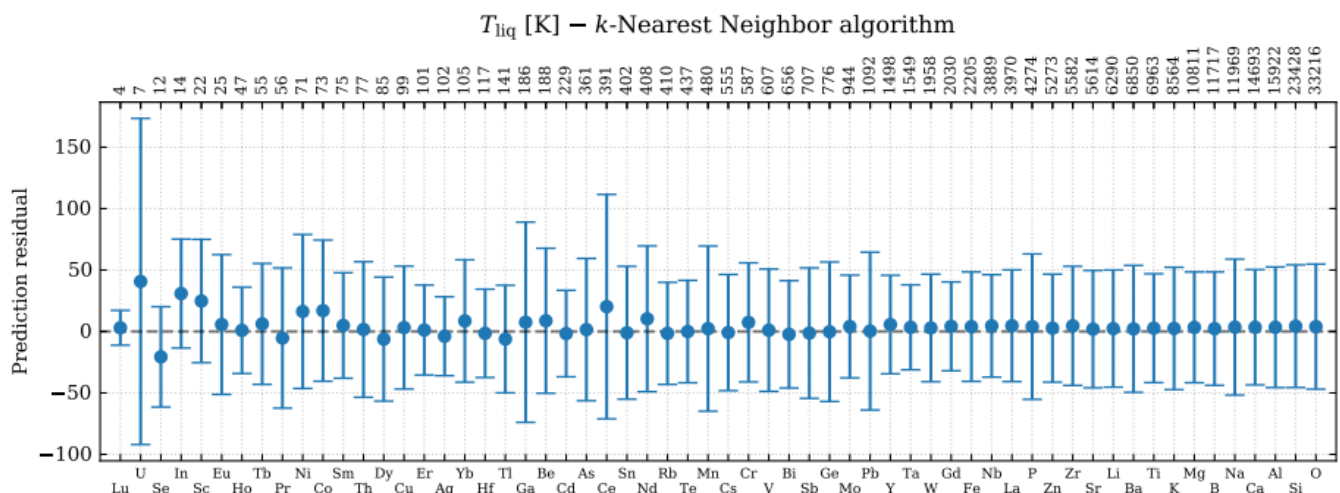

**Figure 6**. Mean and standard deviation of the prediction residual of $T_{liq}$ for each chemical element. The numbers on top are the number of examples containing that element in the dataset. The prediction residual is the difference between the reported and the predicted value.

## 4.6 Refractive index

The refractive index of a transparent material is a dimensionless number that describes how fast light travels through it. This index is formally defined as $n_D = c/v$, where $c$ is the speed of light in vacuum and $v$ is its velocity in this medium. The higher the refractive index, the lower the speed, and the more a light ray is refracted when traveling through the medium. The refractive index depends on the light wavelength; hence it is typically defined as $n_D$, the value for the D line of sodium, 589 nm. This is a property of utmost importance for all types of optical applications.

Table 7 shows the average prediction error for the refractive index of the three ML algorithms. This table shows that, overall, the best predictor for the refractive index is the tuned RF. The other algorithms produced higher predictive errors. The typical relative deviation of 0.8% is minimal and similar to the experimental error for this property. Once more, the tuned version usually performed better than the default version, as one would expect.

**Table 7**: Experimental results for $n_D$. The full dataset contains approximately 45,000 different glass compositions.

| Metric | Cart | | k-NN | | RF | |
|---|---|---|---|---|---|---|
| | Default | Tuning | Default | Tuning | Default | Tuning |
| RD | 1.05 ± 0.02 | 1.13 ± 0.05 | 0.87 ± 0.02 | 0.80 ± 0.04 | 0.76 ± 0.01 | 0.76 ± 0.02 |
| $R^2$ | 0.944 ± 0.006 | 0.944 ± 0.006 | 0.961 ± 0.005 | 0.965 ± 0.005 | 0.969 ± 0.004 | 0.974 ± 0.003 |
| RMSE | 0.044 ± 0.003 | 0.044 ± 0.002 | 0.037 ± 0.003 | 0.035 ± 0.003 | 0.032 ± 0.002 | 0.031 ± 0.002 |
| RRMSE | 0.24 ± 0.01 | 0.24 ± 0.01 | 0.20 ± 0.01 | 0.19 ± 0.01 | 0.18 ± 0.01 | 0.16 ± 0.01 |

The box plots for each ML algorithm are shown in Figures 7a, b, c, which show that the prediction errors are minimal for low and intermediate values of $n_D$, but upsurge with the increasing refractive index. Similar to what we obtained for $T_g$, the best performing algorithm for low and high refractive indexes was CART.

Even with the best algorithm, the prediction errors for high refractive index values reach approximately 10%. Therefore, care must be taken when predicting novel glassmaking compositions with very high values of $n_D$.

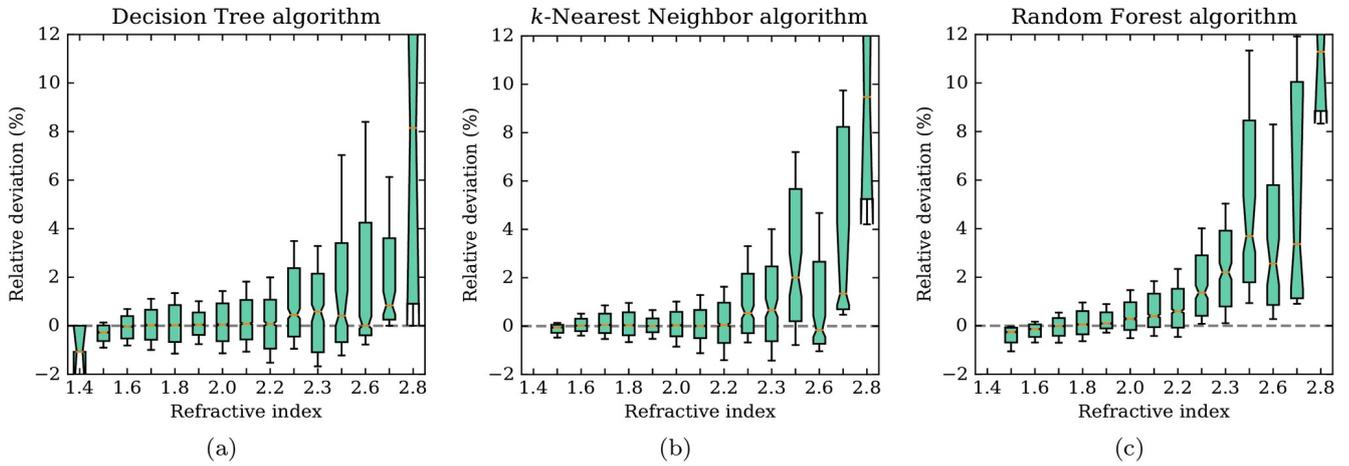

Figure 7. Boxplot of relative deviation (residual of prediction divided by the reported value) of $n_D$ for the tuned models.

Finally, Figure 8 shows the mean and the standard deviation of the prediction residuals of $n_D$ for each chemical element. The model used to build this plot was the one with the lowest RD, the tuned RF. The behavior observed in this plot is reasonably different from that of Figures 4 and 6. The elements with the highest error in prediction are molybdenum, copper, and silver, which are all three transition metals.

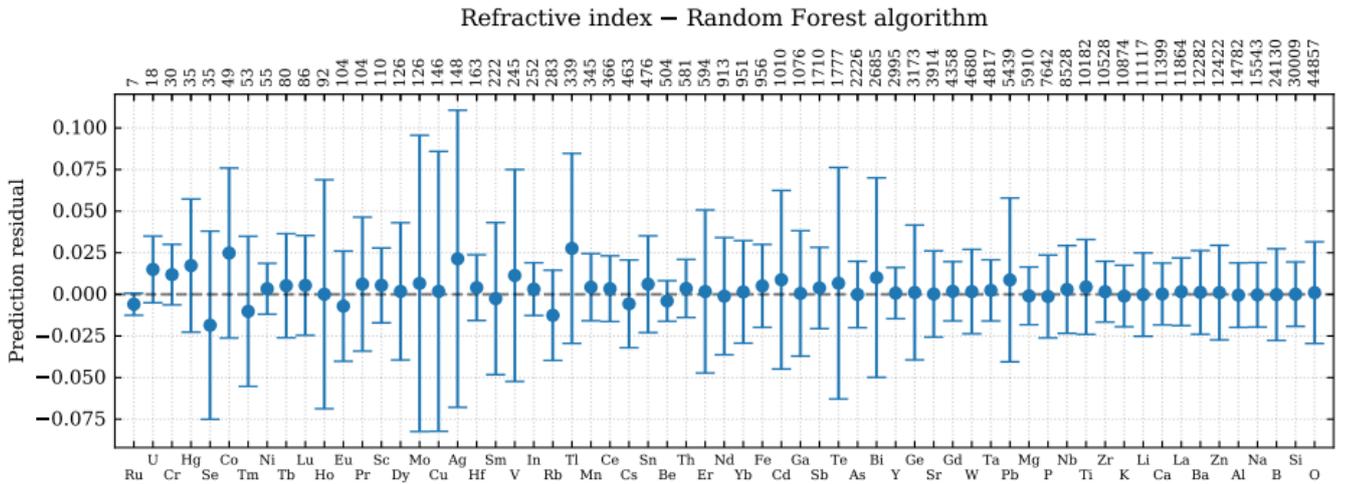

Figure 8. Mean and standard deviation of the prediction residual of $n_D$ for each chemical element. The numbers on top are the number of examples containing that element in the dataset. The prediction residual is the difference between the reported and the predicted value.

Figure 8 shows that the mean prediction residual fluctuates considerably for glasses containing elements with a low number of samples. As the number of samples increases, this mean converges to zero. For glass with chemical elements that have more than 340 samples, the mean prediction residual is systematically lower than 0.015.

## 4.7 Abbe number

The Abbe number ($v_D$) is a measure of dispersion and shows how much the refractive index changes with the wavelength of light. It is defined by Eq. (1), where $n_C$, $n_D$, and $n_F$ are the refractive indices of the material at the wavelengths of 656.3, 589.3, and 486.1 nm, respectively.

$$v_D = \frac{n_D - 1}{n_F - n_C} \qquad (1)$$

Therefore, the smaller the value of $v_D$, the more sensitive the refractive index is to the wavelength. This is a property of paramount importance, e.g., for minimizing chromatic aberration in optical design, for which glasses with a high Abbe number (low dispersion) are required.

**Table 8**: Experimental results for $v_D$. The full dataset contains about 22,000 different glass compositions.

| Metric | Cart | | k-NN | | RF | |
|---|---|---|---|---|---|---|
| | Default | Tuning | Default | Tuning | Default | Tuning |
| RD | 3.7±0.2 | 3.7 ± 0.1 | 3.9 ± 0.1 | 3.6 ± 0.2 | 2.7 ± 0.1 | 2.7 ± 0.1 |
| R² | 0.959 ± 0.005 | 0.959 ± 0.005 | 0.955 ± 0.004 | 0.960 ± 0.004 | 0.977 ± 0.004 | 0.977 ± 0.004 |
| RMSE | 2.8 ± 0.2 | 2.8± 0.2 | 3.0 ± 0.1 | 2.8 ± 0.2 | 2.1 ± 0.2 | 2.1 ± 0.2 |
| RRMSE | 0.20 ± 0.01 | 0.20 ± 0.01 | 0.213 ± 0.009 | 0.20 ± 0.01 | 0.15 ± 0.01 | 0.15 ± 0.01 |

Table 8 shows that, overall, the top predictor for the Abbe number is RF. The two other ML algorithms induced regressors with lower predictive performance. The relative deviation of our best model is 2.7%, which is small and close to errors found in experimental measurements. For this property, no improvement was obtained by tuning the hyperparameters.

The box plots for each algorithm are shown in Figures 9a, b, and c. They show that the prediction errors are much higher for low values of the Abbe number, but they are only slightly higher for the highest values of the Abbe number.

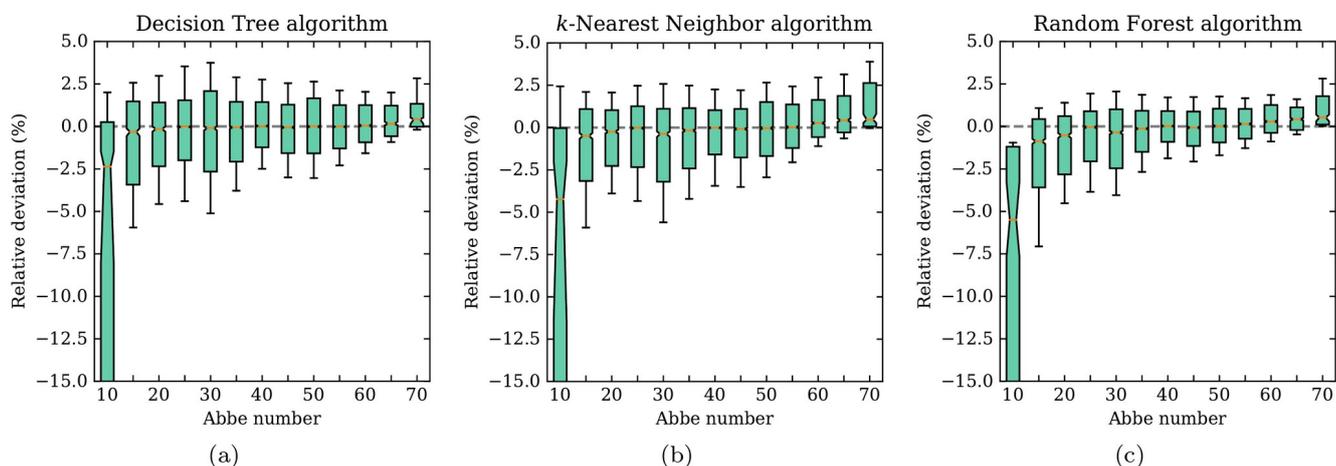

(a) (b) (c)

**Figure 9**. Boxplot of relative deviation (residual of prediction divided by the reported value) of the Abbe number for the tuned models.

Finally, Figure 10 shows the mean and the standard deviation of the prediction residuals of the Abbe number for each chemical element. The model used to build this plot was the one with the lowest RD, the tuned RF. There is a clear trend of having a smaller prediction uncertainty as the number of examples increases (going left to right in Figure 10). In particular, selenium, copper, and vanadium presented the highest standard deviation of the residuals, probably due to the small number of examples with these elements in the dataset (18 at most). This result supports the operation of removing glasses containing chemical elements that rarely appear in the dataset.

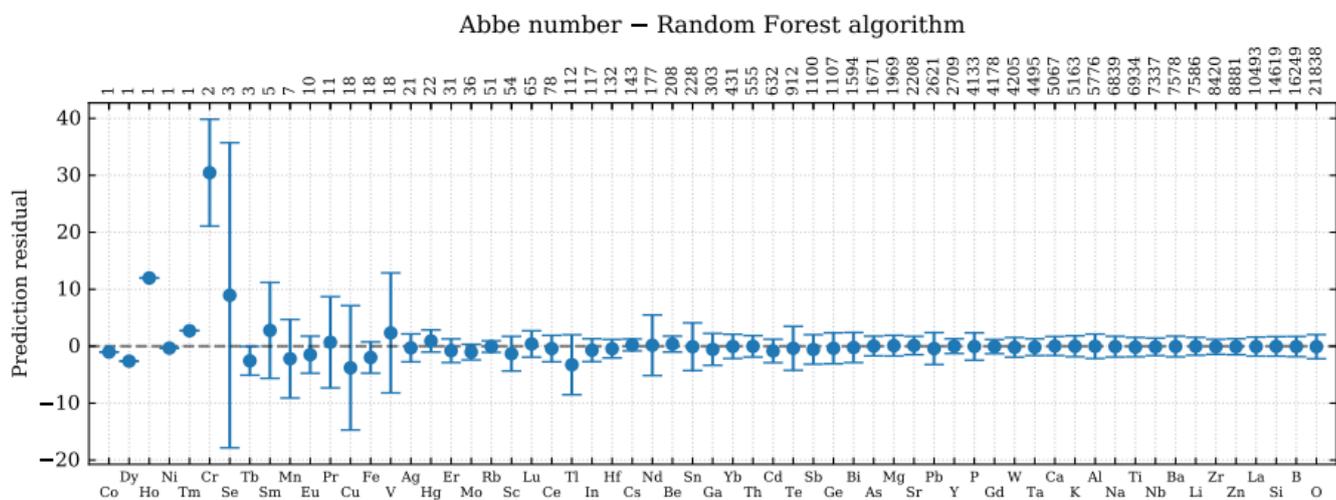

**Figure 10.** Mean and standard deviation of the prediction residual of the Abbe number for each chemical element. The numbers between parenthesis are the number of examples containing that element in the dataset. The prediction residual is the difference between the reported and the predicted value.

Figure 10 shows that the predicted mean fluctuates considerably but consistently approaches the experimental value, and the error bar significantly decreases for a number of examples $N > 20$.

## 4.8 Young's modulus

The Young's modulus (also known as elastic modulus) is the coefficient of the stress-strain curve in the elastic deformation regime of any material. It is a measure of the intrinsic stiffness of a material. It has a crucial bearing on most mechanical applications of all glass types, which are elastic materials and do not show any significant plastic deformation in typical applications.

Table 9 shows that, overall, the top predictors for $E$ were RF and k-NN, both after tuning. The CART algorithm presented the highest predictive errors. The relative deviation of 6% is larger than the typical experimental error obtained in laboratories for this property and much larger than the average RD prediction errors for the other 5 properties (1.0 – 2.5%).

**Table 9**: Experimental results for Young's modulus. The full dataset contains about 13,000 different glass compositions.

| Metric | Cart | | k-NN | | RF | |
|---|---|---|---|---|---|---|
| | Default | Tuning | Default | Tuning | Default | Tuning |
| RD | 6.8 ± 0.5 | 7.0 ± 0.5 | 5.8 ± 0.5 | 5.2 ± 0.4 | 5.6 ± 0.5 | 5.4 ± 0.4 |
| $R^2$ | 0.83 ± 0.03 | 0.83 ± 0.03 | 0.89 ± 0.02 | 0.90 ± 0.02 | 0.90 ± 0.02 | 0.92 ± 0.02 |
| RMSE | 8.6 ± 0.8 | 8.6 ± 0.8 | 6.7 ± 0.5 | 6.4 ± 0.6 | 6.4 ± 0.7 | 6.0 ± 0.6 |
| RRMSE | 0.42 ± 0.04 | 0.42 ± 0.04 | 0.33 ± 0.02 | 0.31 ± 0.03 | 0.31 ± 0.04 | 0.29 ± 0.03 |

The boxplots for each algorithm are shown in Figure 11a, b, and c. These figures show that the prediction errors are significantly larger for the smallest and the largest values of $E$ and that the best performer for extreme values is CART.

Even with the best algorithm, the error for the extreme values is very high. This result indicates that great care must be taken when using these models to predict novel compositions with very low or very high $E$ values. The higher errors in the prediction of $E$ than all the other properties are likely because this dataset contained "only" 13,000 examples, versus 22,000 to 51,000 for the other properties.

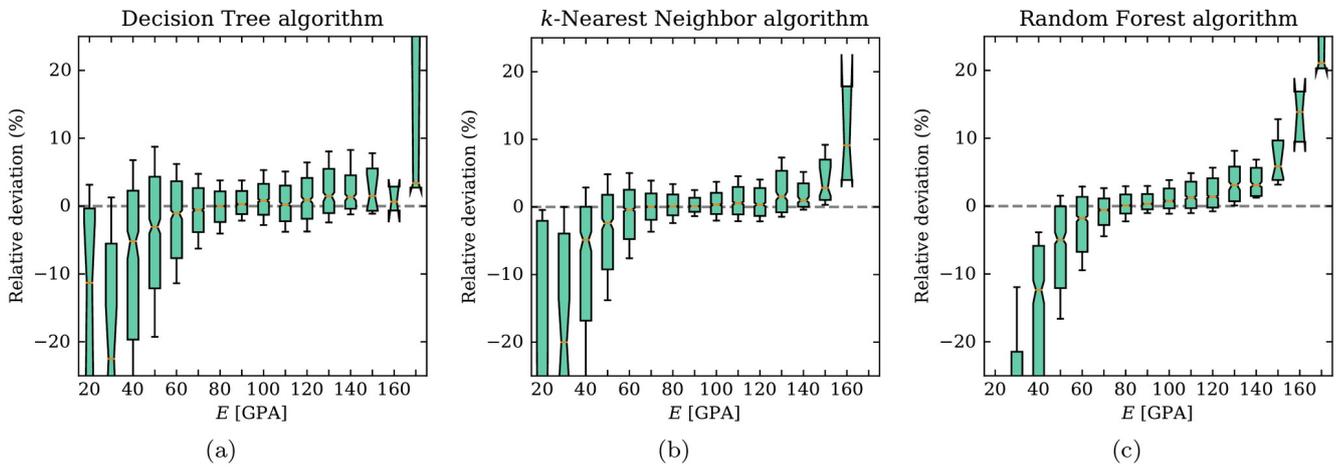

**Figure 11.** Boxplot of relative deviation (residual of prediction divided by the reported value) of Young's modulus for the tuned models.

Finally, Figure 12 shows the mean and standard deviation of the prediction residuals of elastic modulus for each chemical element. The model used to build this plot was the one with the lowest RD, the tuned k-NN. Once more, glasses with chemical elements that are poorly represented had the worst results of prediction. Besides, some glasses with a reasonable number of examples, such as vanadium and bismuth, showed a higher uncertainty in prediction. This figure shows that the predicted mean fluctuates much less and approaches the experimental value for the number of examples N > 30.

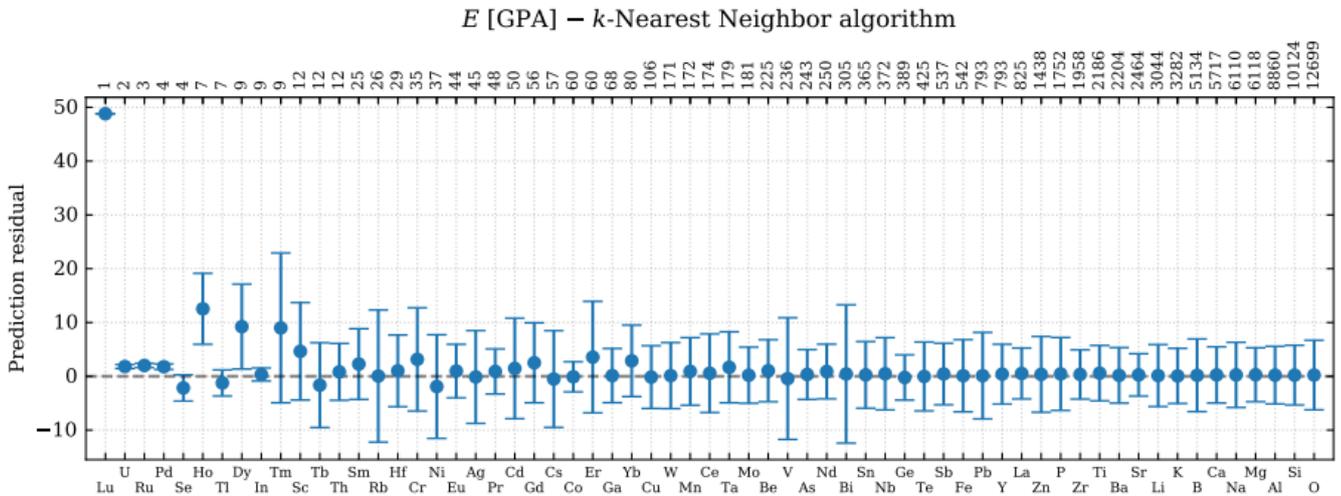

**Figure 12.** Mean and standard deviation of the prediction residual of Young's modulus for each chemical element. The numbers between parenthesis are the number of examples containing that element in the dataset. The prediction residual is the difference between the reported and the predicted value.

## 4.9 Coefficient of thermal expansion

The linear coefficient of thermal expansion is related to the change of size of a material relative to a change in temperature. It is an essential thermal property for designing glasses that work at higher temperatures, such as those used for sealing solid oxide fuel cells.

Table 10 shows that the top predictor for $\log_{10}(CTE)$ was RF, tuned (Tuning) and not tuned (Default), followed by k-NN, tuned and not tuned, respectively. The CART algorithm presented the worst predictive performance values.

**Table 10**: Experimental results for $\log_{10}(CTE)$. The full dataset contains about 51,000 different glass compositions.

| Metric | Cart | | k-NN | | RF | |
|---|---|---|---|---|---|---|
| | Default | Tuning | Default | Tuning | Default | Tuning |
| RD | 0.76 ± 0.02 | 0.80 ± 0.02 | 0.62 ± 0.01 | 0.59 ± 0.02 | 0.57 ± 0.01 | 0.538 ± 0.009 |
| $R^2$ | 0.881 ± 0.006 | 0.883 ± 0.007 | 0.928 ± 0.004 | 0.931 ± 0.006 | 0.934 ± 0.004 | 0.943 ± 0.005 |
| RMSE | 0.070 ± 0.003 | 0.069 ± 0.003 | 0.054 ± 0.002 | 0.052 ± 0.002 | 0.052 ± 0.002 | 0.048 ± 0.002 |
| RRMSE | 0.350 ± 0.010 | 0.34 ± 0.01 | 0.269 ± 0.008 | 0.26 ± 0.01 | 0.258 ± 0.007 | 0.241 ± 0.009 |

Figure 13 illustrates by boxplots the predictive performance obtained by each algorithm, labeled as a, b, and c. According to these figures, the predictive performance obtained by all algorithms was reduced for the extreme $\log_{10}(CTE)$ values. As in the previous results, CART presented the best performance for these values. A relevant observation is that the highest prediction residuals of Figure 13 are *not* located in the extremes, which is different from the results for all other properties studied in this work. Figure 2f shows that the minimum of the distribution is not located at the extremes and that the distribution of $\log_{10}(CTE)$ values is quite different from the other properties, having the lowest skewness and the highest kurtosis. We attribute the distinctive nature of Figure 13 to these particularities of the $\log_{10}(CTE)$ distribution.

Another relevant distinction of the $\log_{10}(CTE)$ dataset, when compared with the dataset of the other properties studied, is how SciGlass aggregates these values. As expected, the thermal expansion coefficient is a property that depends on the temperature of measurement. However, it is expected that changes in CTE below the glass transition temperature are relatively small, if not negligible. This is why the SciGlass database collects and groups CTE data measured below $T_g$ under the same dataset, which was used to induce the models in this work. This grouping strategy could, in principle, also be related to the unusual nature of Figure 13.

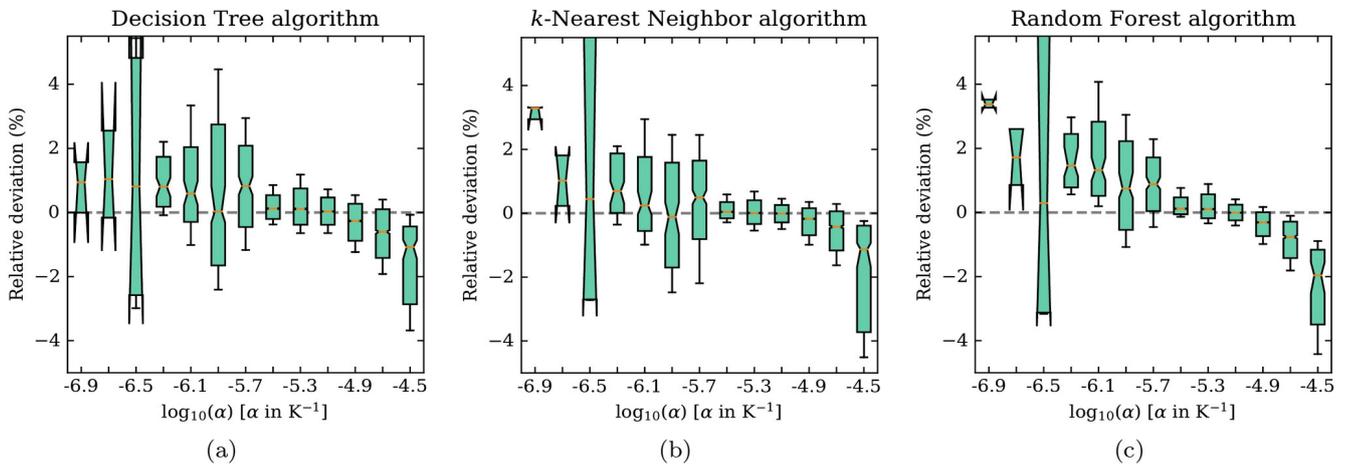

(a)  (b)  (c)

**Figure 13**. Boxplot of relative deviation (residual of prediction divided by the reported value) of $\log_{10}(\text{CTE})$ for the tuned models.

Finally, Figure 14 shows the mean and the standard deviation of the prediction residuals of $\log_{10}(\text{CTE})$ for each chemical element. The model used to build this plot was the one with the lowest RD, the tuned RF. The element with the highest residual standard deviation is also the one with fewer examples, palladium. In this figure, it is relevant to note that the predicted mean fluctuates but consistently approaches the experimental value for a number of examples $N > 500$.

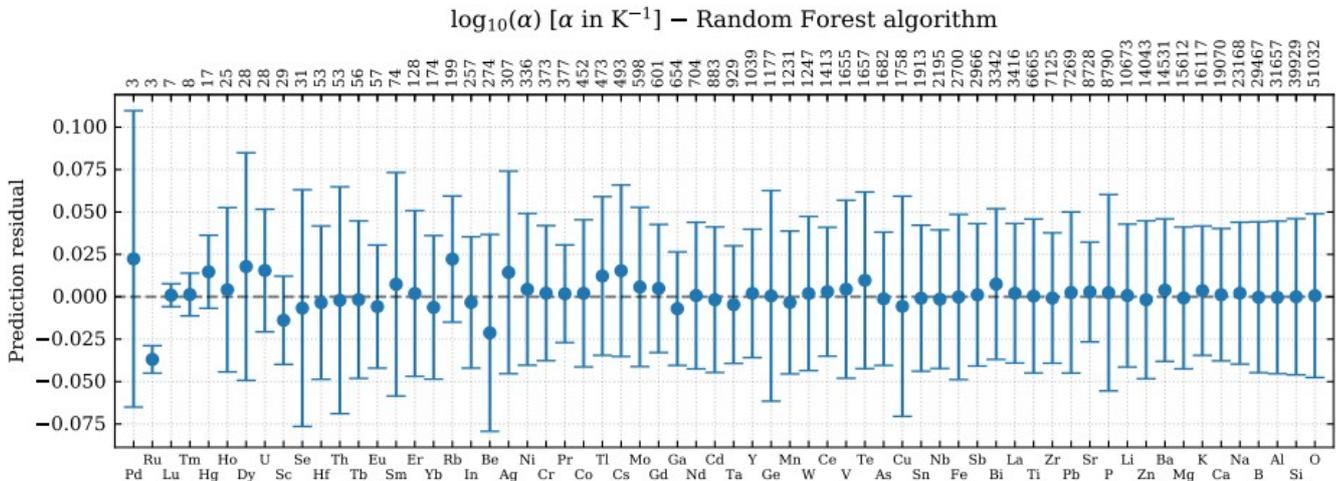

**Figure 14.** Mean and standard deviation of the prediction residual of $\log_{10}(\text{CTE})$ for each chemical element. The numbers between parenthesis are the number of examples containing that element in the dataset. The prediction residual is the difference between the reported and the predicted value.

## 4.10 Overall analysis of the results

While no extreme differences were found, the predicting capabilities of the three ML algorithms used here do differ. The RF and k-NN consistently lead to smaller prediction errors than CART. The RF algorithm generally yielded the best results, closely followed by k-NN. The only exception was for the liquidus temperature, where k-NN performed slightly better. This k-NN result is interesting in itself, as it supports the idea that more features (physical, chemical, and structural) may be necessary to improve the prediction of some properties. In this particular case, even though we are not explicitly giving these features to the algorithms, it is somewhat embedded in the procedure of looking for similar neighbors in the composition domain, which is the principle of the k-NN algorithm. Briefly, the composition of the glasses in the training dataset can be thought of as points in the composition hyperspace, with each chemical element acting as the coordinates of these points. When making a new prediction, the k-NN checks the property of the compositions in its internal database (which is the training dataset) that are closest (e.g., smaller Euclidean distance) to the composition being predicted. Similar compositions are expected to have similar chemistry. It should also be recalled that the liquidus is not a glass property, instead it is a thermodynamic property of crystalline structures, and as such, it depends much more on the atomic structure than the other properties considered in this work.

Another relevant observation is regarding the prediction in the extreme values of the domain. By only looking at the metrics, we lose this information and, as we have shown here, the algorithm with the best average metrics may not be the one that best predicts the extreme values. We argue that the choice of ML algorithm will depend on the uses of the prediction model: if one wants a general-purpose model for interpolation, then choose the one with the best average predictive performance; on the other hand, if the intention is to find new glasses outside of the envelope with extreme properties, then this model may not be the most suitable.

As future work, it would also be interesting to apply more sophisticated data cleaning techniques to the dataset to identify and possibly remove outliers; this is one of the 21 challenges of this field [27]. We also want to investigate alternatives to mitigate the poor predictive performance on extreme values.

## 4.11 SHAP analysis

Figure 15 shows all the beeswarm plots obtained by the SHAP analysis described in the materials and methods section. These plots can give valuable insights for designing new glasses, as discussed in the following paragraphs.

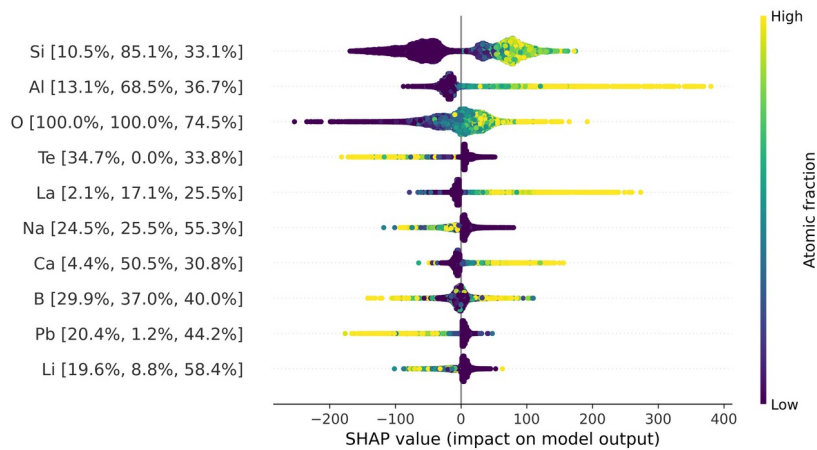

(a) $T_\text{g}$, the base value is 777 K.

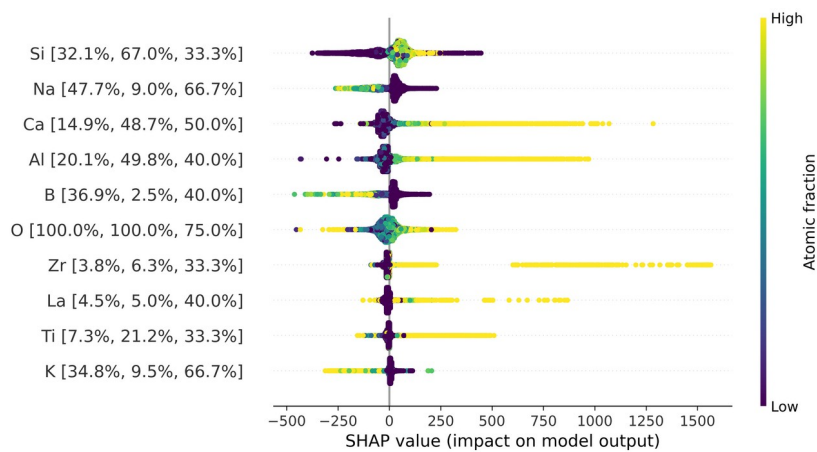

(b) $T_\text{liq}$, the base value is 1410 K.

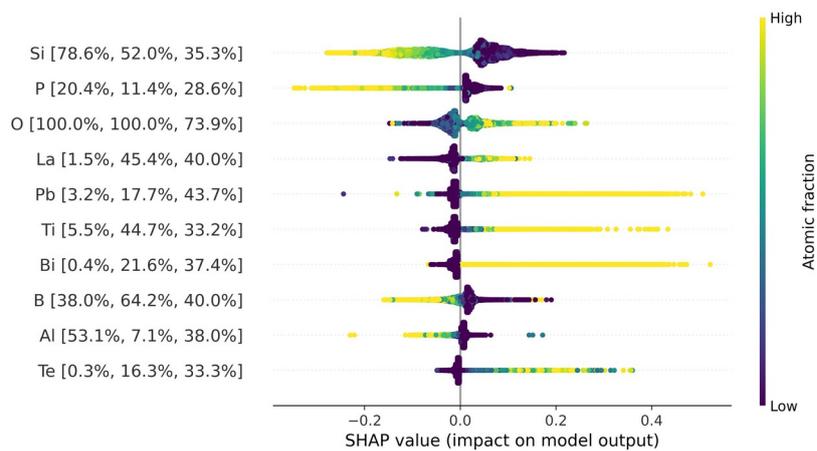

(c) $n_\text{D}$, the base value is 1.69.

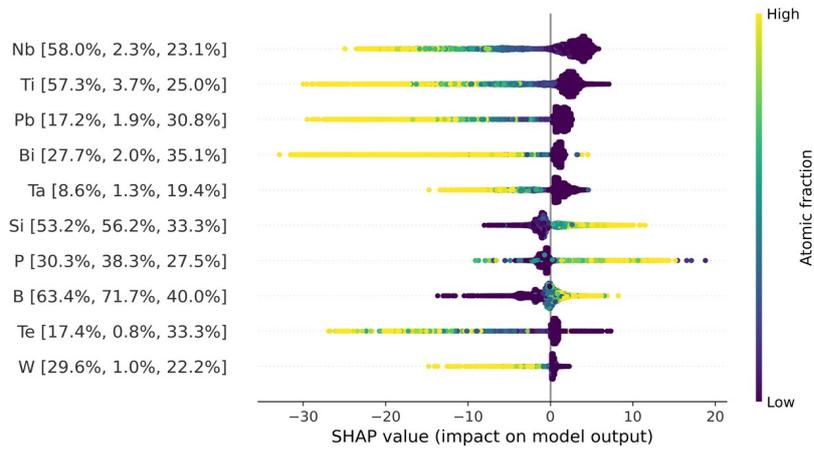

(d) $v_D$, the base value is 43.0.

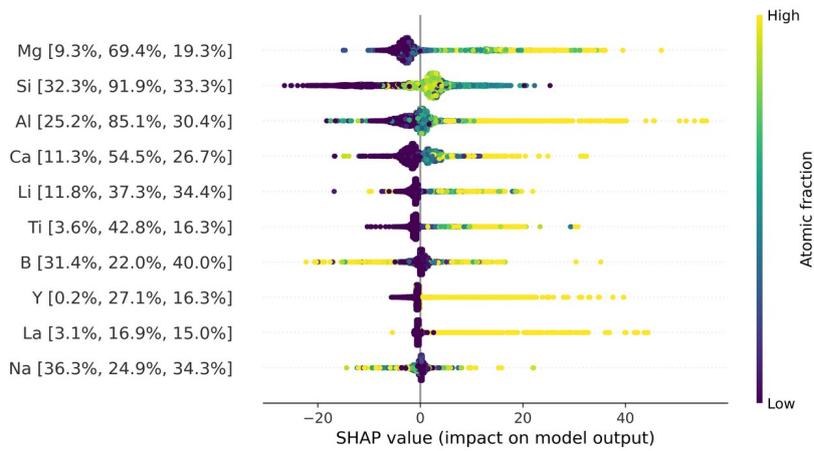

(e) $E$, the base value is 77.0 GPa.

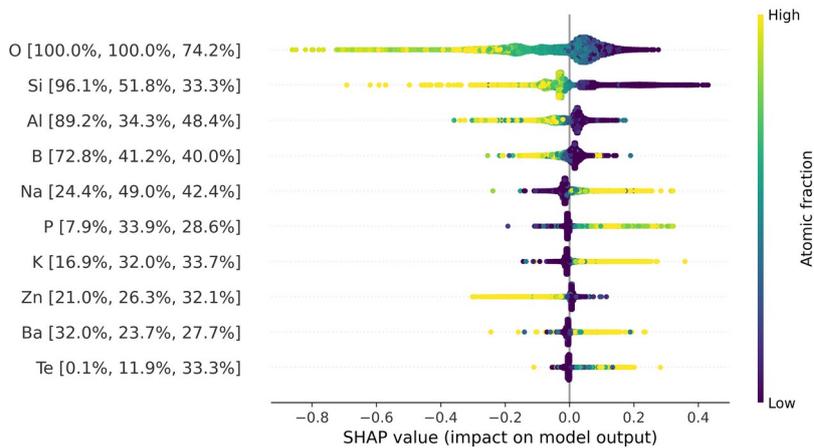

(f) $\log_{10}(CTE)$, the base value is $-5.12$.

**Figure 15.** Beeswarm plot of the SHAP values of the six studied properties. The left label shows the ten most important elements identified by the SHAP analysis in decrescent order of importance. The three numbers inside the brackets are the fraction of glasses in the low range of the property (lower than the 20% percentile) having the respective chemical element; the fraction of glasses in the high range of the property

(higher than the 80% percentile) having the respective chemical element; and the maximum atomic fraction of the respective chemical element in the dataset.

For the *glass transition temperature*, Figure 15a shows that silicon, aluminum, lanthanum, and calcium increases this property with the increase of their atomic fraction in the glass, the first being a glass-former and the other two intermediate elements. Aluminum and lanthanum can lead to the highest increase in $T_g$, reaching up to 400 K for aluminum. On the other hand, tellurium, sodium, lead, and lithium all contribute to reducing $T_g$. The well-known boron anomaly can be seen in this figure, as increasing the boron content can decrease or increase $T_g$. Glasses having a higher fraction of oxygen (that is, those made with higher valence elements) tend to have higher $T_g$ than those with a lower fraction of oxygen.

For the *liquidus temperature*, Figure 15b shows that calcium, aluminum, zirconium, lanthanum, and titanium can increase this property, from which zirconium shows the most substantial effect. Sodium, boron, and potassium have opposite behavior, reducing $T_{liq}$. No clear trend is found for silicon and oxygen.

For *refractive index*, Figure 15c shows that silicon, phosphorus, boron, and aluminum decrease this property. This is quite interesting as all of them, except the last, are glass-formers. As glass-formers are necessary to make commercial oxide glasses, the only other glass-forming element remaining that increases $n_D$ is tellurium. Lanthanum, lead, titanium, and bismuth are well-known heavy elements that increase $n_D$, and this trend was revealed in this analysis. Similar to the glass transition temperature, glasses with a higher atomic fraction of oxygen also show a higher refractive index.

For the *Abbe number*, Figure 15d shows that niobium, titanium, lead, and bismuth tend to decrease this property by increasing their atomic fraction in the glass. Bismuth can decrease the Abbe number by up to 30, while niobium can decrease it a bit less, approximately 23. The glass-formers silicon, phosphorus, and boron increase the Abbe number, with phosphorus yielding the highest possible increase of approximately 20. Tellurium, on the other hand, is a glass-former that decreases the Abbe number.

For *Young's modulus*, Figure 15e shows that magnesium, aluminum, calcium, lithium, titanium, yttrium, and lanthanum increase this property. Silicon also seems to increase $E$, but the picture is not as clear. Boron clearly shows the boron anomaly again in this plot, similar to what occurs in Figure 15a. Curiously, sodium also seems to have some hybrid effect on $E$ when present in higher amounts in the glass.

Finally, for the base-10 logarithm of the linear *coefficient of thermal expansion*, Figure 15f shows that sodium, phosphorus, potassium, barium, and tellurium increase this property, with tellurium being the only glass-former in this group. Silicon, aluminum, boron, and zinc generally reduce the CTE, with silicon having the most substantial impact on this property. Glasses with a high fraction of oxygen tend to have lower CTE.

Some insights obtained from the SHAP analysis are well-known by the glass community or can be easily observed by simple data analysis; for example, there are many more glasses with a low Abbe number (58%) with niobium than glasses with a high Abbe number (2.3%). This is no surprise; niobium is known for its tendency to reduce this property; however, it is impressive to know from Fig. 15d the *intensity* of such reduction. In fact, the SHAP analysis suggests that in small quantities (up to an atomic fraction of 2%), adding niobium increases the Abbe number with respect to the base value! These insights gained by the SHAP analysis can help both empirical glassmaking (by guiding compositional tuning) and computer-aided inverse design of glasses [26,40–42] by helping restrict the search domain for complex problems.

## 5. Summary and Conclusions

We trained and investigated the predictive performance of three ML algorithms for six glass properties using an extensive dataset of approximately 150,000 oxide glasses. We induced predictive models for glass transition temperature, liquidus temperature, elastic modulus, thermal expansion coefficient, refractive index, and Abbe number using decision tree induction, k-nearest neighbors, and random forest algorithms. Each model was induced with its default set of hyperparameters and with its hyperparameters tuned via a random search. The models induced by RF and k-NN often yielded the best prediction metrics overall but fail to predict extreme values of the properties.

We demonstrate that, apart from the elastic modulus (for which "only" 13,000 compositions were available), predictive models with high performance can be induced for the other five properties, yielding comparable levels of uncertainties as those of the usual data spread. However, care must be taken when using these predictive models for glasses that have properties with extremely low or high values, as the uncertainty in these cases is significantly higher. As expected, glasses containing chemical elements that are poorly represented in the training set often present higher prediction errors. Despite this weakness, we believe the method used here will be beneficial for selecting or designing new glasses having a suitable combination of properties.

SHAP analysis indicated the key elements that increase or decrease the value of any desired property. It also estimates the *maximum* possible increase or decrease of a property. Such insights can help both empirical compositional tuning and computer-aided inverse design of glass formulations.

The six induced models of this work can be used for the inverse task of prescribing chemical compositions. The goal is to use these induced models to recommend the chemical composition that might yield the desired combination of any of these six properties.

**Acknowledgments**

This study was financed by the São Paulo Research Foundation (FAPESP) grants number 2017/12491-0, 2018/07319-6, 2017/06161-7, 2018/14819-5, 2013/07375-0, and 2013/07793-6.

**Competing interest statement**

The authors declare no competing financial or non-financial interests.

# Supplementary Material to "Predicting and interpreting oxide glass properties by machine learning using large datasets"

Daniel R. Cassar[1], Saulo Martiello Mastelini[2], Tiago Botari[2], Edesio Alcobaça[2], André C.P.L.F. de Carvalho[2], Edgar D. Zanotto[1]

[1]Department of Materials Engineering, Federal University of São Carlos, São Carlos, Brazil
[2]Institute of Mathematics and Computer Sciences, University of São Paulo, São Carlos, Brazil

## Hyperparameters Tuned

In this supplementary session we describe the set and range of hyperparameters used during the tuning of the regression algorithms. The information is shown in Table S.1.

**Table S.1**: Hyperparameter search space and best values obtained after tuning for the six studied properties.

Table 1: Hyperparameters tuned for the k-NN algorithm along with their best values found

| Hyperparameter | Description | Range | Best value $\nu$D | nD | $\log_{10}$(CTE) | Tliq | E | Tg |
|---|---|---|---|---|---|---|---|---|
| n_neighbors | Number of neighbors to consider in the algorithm. | [1, 100] | 4 | 5 | 6 | 4 | 4 | 3 |
| weights | Strategy adopted to combine the neighbor information: either "uniform" or "distance". The former option means that simple arithmetic mean will be performed to combine the neighbors' target values; the latter option indicates that the combination will use weights inversely proportional to the distance of the neighbor to the example of interest. | – | distance | distance | distance | distance | distance | distance |

Table 2: Hyperparameters tuned for the CART algorithm along with their best values found

| Hyperparameter | Description | Range | Best value $\nu$D | nD | $\log_{10}$(CTE) | Tliq | E | Tg |
|---|---|---|---|---|---|---|---|---|
| criterion | Criterion used for node splitting. We explored two possible values, which were either "mse" or "friedman_mse". The first option refers to the traditional mean square error (MSE), while the second one uses MSE along with the Friedman's improvement score for potential splits. | – | friedman_mse | friedman_mse | friedman_mse | friedman_mse | mse | mse |
| min_impurity_decrease | The minimum amount of impurity a split must reduce in order to be performed. The term impurity refers to the split criterion used. The impurity reduction calculation is weighted by the number of examples lying in each induced tree branch. | [0, 0.1] | 0.0061 | 0.0012 | 0.0336 | 0.0431 | 0.0024 | 0.0601 |

Table 3: Hyperparameters tuned for the RF algorithm along with their best values found

| Hyperparameter | Description | Range | Best value $\nu$D | nD | $\log_{10}$(CTE) | Tliq (K) | E (GPa) | Tg (K) |
|---|---|---|---|---|---|---|---|---|
| n_estimators | Indicates the number of trees of the ensemble. | [100, 1000] | 885 | 390 | 912 | 948 | 595 | 991 |
| max_features | The maximum number of features to consider at each split attempt. We considered three built-in settings for RF: "auto", "sqrt", and "log2". The first option uses n_features, the second option $\sqrt{n\_features}$, and the last one $\log_2(n\_features)$. | – | auto | sqrt | sqrt | log2 | sqrt | sqrt |

34